%% file: main.tex
\documentclass[fullpage,twocolumn]{article}
\pdfoutput=1

\usepackage{times}
\usepackage{xspace}
\usepackage{cite} % sort citation numbers
\usepackage[hyperindex,breaklinks,colorlinks,
	    linkcolor=blue,citecolor=blue,urlcolor=black]{hyperref}
\usepackage{cleveref}
\usepackage{graphicx}
\usepackage[T1]{fontenc}

\long\def\com#1{}
\newcommand{\xxx}[1]{}
\long\def\arxiv#1#2{#1}			% real version
\newcommand{\ie}{{\em i.e.},\xspace}
\newcommand{\eg}{{\em e.g.},\xspace}

\newcommand{\ml}{$\star$ML\xspace}	% The SGML-derived markup languages

\begin{document}

\title{Matchertext: Towards Verbatim Interlanguage Embedding \\
	\Large{%\texttt{preliminary draft - not yet for redistribution} \\
	~\\
	%Working project repository:
	\url{https://github.com/dedis/matchertext}}}

\author{Bryan Ford \\ EPFL}

\maketitle

\input{abs}

\tableofcontents

\input{intro}

\input{bg}

\input{design}

\input{host}

\input{embed}

% ...
\input{impl}

\input{eval}

\input{rel}

\input{concl}

\bibliographystyle{plain}
\arxiv{
\bibliography{lang,net,sec,soc}
}{
\bibliography{main}
}

\end{document}

%% file: abs.tex
\begin{abstract}
Embedding text in one language within text of another is commonplace
for numerous purposes,
but usually requires tedious and error-prone
``escaping'' transformations on the embedded string.
We propose a simple cross-language syntactic discipline,
\emph{matchertext},
which enables the safe embedding a string in any compliant language
into a string in any other language via simple ``copy-and-paste'' --
in particular with no escaping, obfuscation, or expansion of embedded strings.
We apply this syntactic discipline
to several common and frequently-embedded
language syntaxes such as URIs, HTML, and JavaScript,
exploring the benefits, costs, and compatibility issues
in adopting the proposed matchertext discipline.
One early matchertext-based language is MinML,
a concise but general alternative syntax for writing HTML or XML.
\end{abstract}

%% file: intro.tex
\section{Introduction}
\label{sec:intro}

The need to embed valid strings in one language into valid strings in another
is commonplace throughout programming practice.
Just a few examples include embedding
regular expressions, URIs, SQL queries, or HTML markup
within string constants in general-purpose programming languages;
embedding user-entered web form data or JavaScript variables into SQL queries;
JavaScript code embedded in HTML via the \verb|<script>| element;
and URIs embedded as query strings within other URIs,
such as in a query to a service like the
\href{https://archive.org/web/}{Wayback Machine}.

An equally-ubiquitous issue arising from this practice
is the need to transform the embedded string --
generally by \emph{escaping} certain characters sensitive
to the ``host'' language and syntactic context --
so that host language processors will not misinterpret embedded text
as host-language text.
For example,
the regular expression \verb|"[^"]*"| matches double-quoted strings,
but when embedded in a C-like language must be written
like \verb|re.match("\"[^\"]*\"")|,
escaping all the embedded instances of double-quote characters,
so that the embedded quotes will not prematurely end the string literal.
Accidentally forgetting necessary escaping
is naturally a common source of syntax errors in manual embedding practice.
\emph{Automated} embedding is also common practice, however,
such as accepting an arbitrary user-entered string on a web form
and embedding it into HTML via a scripting language like PHP.
With automated embedding,
forgetting to escape embedded strings properly
has become an endless source of critical security bugs
such as SQL injection~\cite{clarke12sql}.
or cross-site scripting attacks~\cite{fogie07xss}.

In some future evolution
of today's standard programming languages and practices,
could we achieve the ability to embed any valid string
in essentially any language into any other --
\emph{across languages} --
without ever having to escape, or otherwise transform or obfuscate,
the string to be embedded?
Could we make embedding always a simple matter of verbatim ``copy-and-paste''
when done manually,
or a simple matter of concatenation
or filling a ``hole'' in a tempate
when done automatically?
We propose that the answer can and should be \emph{yes} --
though with important challenges, costs, and caveats of course.

We observe that verbatim interlanguage embedding would be achievable if:
(1)
we could standardize across languages
a set of open/close character pairs
we will call \emph{matchers},
such as the parentheses \verb|()|,
square brackets \verb|[]|,
and curly braces \verb|{}|;
and
(2) 
we could impose the universal ``syntactic discipline''
that \emph{matchers must properly match} in nested pairs
throughout any valid string -- without exception --
in any compliant language.
If \emph{plain text} is an unstructured linear sequence of characters
in a character set like ASCII or Unicode,
then we define \emph{matchertext} to be plain text
conforming to the additional syntactic discipline
that ASCII matchers must match.
For example, the strings `\verb|(a{b}c)|' and `\verb|a({'}["])d|'
are valid matchertext,
while the strings
`\verb|(|', `\verb|{a]|', `\verb|[(])|', and `\verb|}{|'
are plain text but are not valid matchertext.

\begin{figure*}[t]
\begin{center}
\begin{small}
\begin{tabular}{ll}
URI syntax:	& \verb|http://search.engine/linksto?site=http%3A%2F%2Fmy.site%2F&results=50| \\
MRI syntax:	& \verb|http[//search.engine/linksto?site=http[//my.site/]&results=50]| \\
\\
URI syntax:	& \verb|http://historical.archive/get?site=http%3A%2F%2Fmy.site%2F&year=1998| \\
MRI syntax:	& \verb|http[//historical.archive/get?site=http[//my.site/]&year=1998]| \\
\end{tabular}
\end{small}
\end{center}
\caption{Example queries containing embedded resource identifiers
	in URI and MRI syntax for comparison.}
\label{fig:search-query}
\end{figure*}

Consider \emph{matchertext resource identifiers} (MRIs),
a matchertext adapation of
uniform resource identifiers (URIs)~\cite{rfc3986} or
internationalized resource identifiers (IRIs)~\cite{rfc3987}.
A URI like \verb|http://my.site/path/|
may always be transformed to or from
equivalent MRI syntax like \verb|http[//my.site/path]|.
An MRI is embeddable verbatim, with no transformation,
into another MRI or into another matchertext-aware language.
Figure~\ref{fig:search-query} shows two example search queries
containing embedded resource identifiers,
contrasting ``copy-and-paste'' embeddable MRI syntax
with traditional URI syntax where the sensitive
colon (\verb|:|) and slash (\verb|/|) characters
must be escaped as \verb|%3A| and \verb|%2F%|, respectively.

\begin{figure*}[t]
\begin{center}
\includegraphics[width=0.95\textwidth]{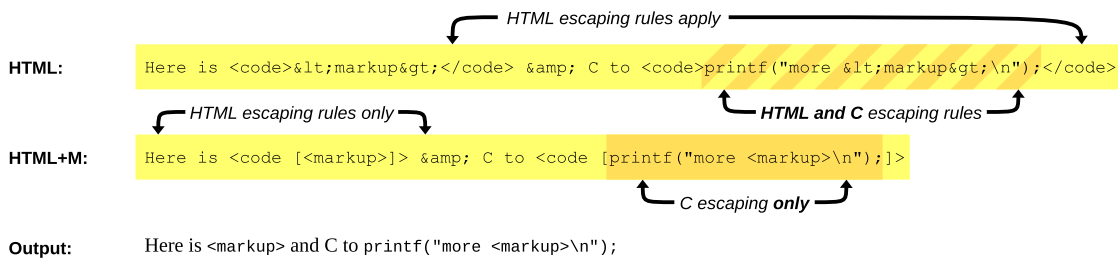}
\end{center}
\caption{Illustration of how different languages' escaping rules
	combine to increase syntactic complexity
	in traditional embedding practice,
	while in matchertext only one language's escaping rules
	are ever active at a given text position.}
\label{fig:territorial-integrity}
\end{figure*}

Adopting the matchertext discipline
does not eliminate the need for character escape sequences:
in fact it can slightly increase the ``escaping obligations''
within a language, as discussed below.
But matchertext enables languages to preserve
the ``territorial integrity'' of their escaping and other syntactic rules,
ensuring that developers need to think about
\emph{only one language's rules at a time}
at any given text position,
even in a string composed from multiple languages.
Thus, matchertext arguably reduces the cognitive load
of writing (or reading) cross-language embedded code.
Figure~\ref{fig:territorial-integrity} illustrates this difference
with a simple example with C code embedded in HTML
including the use of escape codes in both languages.

Existing languages could be adapted incrementally
to support and leverage matchertext.
C-like languages, for example,
might adopt a new escape sequence like \verb|\[|$m$\verb|]|
to embed an arbitrary matchertext $m$
into a quoted string or character literal.
All characters including quotes and newlines
are allowed within $m$,
provided only that matchers match.
Thus,
%\verb|'\[']'| would be equivalent to \verb|'\''|, and
the earlier string-matching regular expression
\verb|"[^"]*"|
could be embedded into a string literal
like \verb|re.match("\["[^"]*"]")|.
With this extension, we use an escape sequence
to delimit the entire embedded matchertext,
but we no longer need to add escape sequences
\emph{within} the embedded regular expression.

Many languages already support \emph{raw string literals}
in which escape sequences are disabled:
\eg backtick strings like \verb|`\`| in Go.
A particular terminating character or sequence
must still be forbidden throughout the embedded text,
however -- in this case the backtick (\verb|`|).
Verbatim copy-and-paste embedding remains unsafe
without carefully checking the embedding text
for the forbidden host-language delimeters,
and changing the delimeters or awkwardly rewriting the embedded text
(\eg by concatenating multiple strings in different quoting styles).
Matchertext, in contrast, forbids \emph{no} characters or sequences
in embedded strings provided only that matchers match.

It is already feasible to write code in existing languages
that is also valid matchertext,
with a bit of care.
Most languages use the matchers in structurally paired forms anyway,
as in expressions like \verb|a*(b+c)|,
lists like \verb|[a,b,c]|,
or maps like \verb|{a:1, b:"hi"}|.
The challenge is mainly in handling the exception cases
where unmatched matchers may commonly appear.

\begin{table*}[t]
\begin{center}
\begin{footnotesize}
\begin{tabular}{ll||ll|ll|ll||ll|ll}
&		& \multicolumn{6}{|l||}{Escapes for C-like languages}
		& \multicolumn{4}{|l}{Escapes for SGML-derived languages} \\
&		& \multicolumn{2}{|l|}{octal escapes}
		& \multicolumn{2}{|l|}{hex escapes}
		& \multicolumn{2}{|l||}{\bf matchers (new)}
		& \multicolumn{2}{|l|}{entity names}
%		& \multicolumn{2}{|l|}{hex}
		& \multicolumn{2}{|l}{\bf matchers (new)}		\\
		&
		& open			& close
		& open			& close
		& open			& close
		& open			& close
		& open			& close		\\
\hline
Parentheses	& \verb|()|
		& \verb|\050|		& \verb|\051|
		& \verb|\x28|		& \verb|\x29|
		& \verb|\o()|		& \verb|\c()|
		& \verb|&lpar;|		& \verb|&rpar;|
%		& \verb|&#x28;|		& \verb|&#x29;|
		& \verb|&(<);|		& \verb|&(>);|	\\
Square brackets	& \verb|[]|
		& \verb|\133|		& \verb|\135|
		& \verb|\x5B|		& \verb|\x5D|
		& \verb|\o[]|		& \verb|\c[]|
		& \verb|&lbrack;|	& \verb|&rbrack;|
		& \verb|&[<];|		& \verb|&[>];|	\\
		
Curly braces	& \verb|{}|
		& \verb|\173|		& \verb|\175|
		& \verb|\x7B|		& \verb|\x7D|
		& \verb|\o{}|		& \verb|\c{}|
		& \verb|&lbrace;|	& \verb|&rbrace;|
		& \verb|&{<};|		& \verb|&{>};|	\\

\end{tabular}
\end{footnotesize}
\end{center}
\caption{Potential alternatives in C- and SGML-derived languages
	to escape unmatched matchers in matchertext.}
\label{tab:unmatched-matchers}
\end{table*}

One traditional habit we must awkwardly unlearn in matchertext,
unfortunately,
is using unmatched matchers in quoted strings
to parse or print structured text.
Clauses like \verb|printf("{")| or \verb|case "]"|
are not valid matchertext.
We must therefore escape such unmatched matchers in literals,
as in \verb|printf("\x7B")| or \verb|case "\x5D"| for example.
Backward-compatible language extensions might ease this pain,
however, 
with new escape sequences that include \emph{both} matchers of a pair
but select only the opener or closer.
The proposed new escape \verb|\o()| represents a literal open parenthesis,
for example,
while \verb|\c[]| represents a close bracket.
Table~\ref{tab:unmatched-matchers} summarizes a few existing and proposed
alternatives for escaping unmatched matchers
in both C-like languages and SGML-derived languages like HTML.

In the rest of this paper, 
we develop more deeply the design and rationale for the matchertext discipline,
then explore how a number of common languages of varying types
might be incrementally adapted to support and leverage
the matchertext discipline effectively.

This work is in an early exploration and experimentation phase,
so the evaluation is currently a placeholder,
serving as a preliminary map for the ways in which
we \emph{would like to} evaluate matchertext
and its use in practical languages.
Some key questions we would like to answer include:
how common (and how painful) is the need for escaping
in the most common cross-language embedding scenarios?
How extensively would large existing repositories of code or data
need to be modified in order to convert them to matchertext?
How does matchertext affect the usability to users or developers
of common constructs in common embedding use-cases,
such as synthesizing or editing HTTP or SQL queries?
How does matchertext affect the frequency of syntax-related bugs --
especially those potentially leading to security vulnerabilities --
in code from typical developers?

One early pragmatic experiment with matchertext, however,
is MinML~\cite{ford22minml},
a concise but general-purpose alternative syntax
for SGML-derived markup languages such as HTML and XML.
MinML uses matcher characters for structure
instead of matching pairs of start and end tags:
\eg \verb|em[emphasis]| rather than \verb|<em>emphasis</em>|,
and \verb|"[quotation]| instead of \verb|&ldquo;quotation&rdquo;|.
MinML supports escapeless embedding of matchertext in other languages
via a sequence like \verb|+[matchertext]|,
as well as unmatched matcher escapes
like those in \cref{tab:unmatched-matchers}.
An \href{https://github.com/bford/hugo}{experimental extention}
to the \href{https://gohugo.io}{Hugo} static website generator
already allows web authoring in MinML.

\xxx{
Caveats:
- does not eliminate other needs for escaping (just escaping for embedding).
- does not eliminate the need for correct input validation
in automated embedding
(e.g., checking that the input is matchertext
or transforming it into matchertext)),
and thus cannot be expected to eliminate all syntax-related security bugs
of the SQL injection variety.
Mistakes will still happen.
However, we may hope that the consistent and rigorous application
of the matchertext discipline
might significantly increase the robustness of embedding practices
and hence decrease the frequency of such bugs.
XXX to be tested empirically.
}

\xxx{ paper roadmap}

\subsection*{An open research project}

This draft represents a ``work-in-progress'' snapshot
of an experimental open research project.
Anyone with adequate interest, skills, and motivation
is welcome to contribute to this research project,
and potentially become a co-author upon making a substantial contribution.
(Smaller contributions will receive acknowledgments in the final paper.)
To propose a contribution, please use Pull Requests (PRs)
on the project's \href{https://github.com/dedis/matchertext}{GitHub repository}.
We do not have time and cannot promise to answer all E-mails,
or provide detailed guidance,
before an interested potential contributor has proactively
created and submitted some significant and well-considered contribution.

%% file: bg.tex
\section{Background: needs and pitfalls of interlanguage embedding}
\label{sec:bg}

The practice of
embedding strings from one language into another
is ubiquitous --
as is the pain of having to ``escape'' embedded strings
to protect them from misinterpretation by the host language processor.
This section briefly explores a few of these common existing practices
and the syntactic composition problems they create.

\subsection{Special-purpose languages}

Many special-purpose language syntaxes exist
almost solely for embedded use in other syntactic contexts.
A few particularly common examples of such ``little languages''
include regular expressions (REs),
uniform resource identifiers (URIs),
JavaScript Object Notation (JSON),
and Structured Query Language (SQL).

Perhaps the most classic ``little language'' is regular expression (RE) syntax,
commonly used for pattern-based searches and replacements in freeform text.
RE syntax traditionally uses many punctuation characters for special purposes,
but must also allow arbitrary embedded text to be matched literally.
RE syntax therefore makes heavy use of ``backslash-escaping'' in literal text:
\eg the RE \verb|.*|
matches any number of arbitrary characters other than newlines,
while one must write \verb|\.\*| to match the literal string `\verb|.*|'.
Because REs themselves are often embedded in another language --
frequently a C-inspired language that \emph{also} uses backslash-escaping
in string literals --
we must further backslash-escape the backslashes in RE syntax.
A C string literal containing the latter RE above
is written \verb|"\\.\\*"|,
for example.
To match the double-backslash \verb|\\| that begins
a \href{https://en.wikipedia.org/wiki/Path_(computing)#Universal_Naming_Convention}{UNC name} (as in \verb|\\host\path\file|),
the backslashes must be doubled to become \verb|\\\\| as an RE,
then doubled again to become \verb|"\\\\\\\\"| as a C sring literal
containing that RE.
This confusing multi-level explosion of backslash escapes
has been aptly dubbed
\href{https://en.wikipedia.org/wiki/Leaning_toothpick_syndrome}{\emph{leaning toothpick syndrome}}.

\subsection{The complexity of multi-level escaping}

As \cref{fig:territorial-integrity} illustrates,
each additional level of embedding in traditional syntax
adds a set of escaping requirements that the writer (or reader) of code
must carefully consider and apply correctly --
\emph{in the correct order} --
in order to ``guide'' the embedded text
through the levels of host syntax that the text is embedded in.
REs and C string literals each require
a \emph{different} set of punctuation characters to be backslash-escaped,
further increasing the cognitive load when embedding.
\href{https://www.pcre.org/original/doc/html/pcrepattern.html}{PCRE syntax}
helpfully promises that a backslash followed by any non-letter
always ``takes away any special meaning that character may have'',
so one may fall back on just \emph{escaping all punctuation}
instead of remembering which characters must actually be escaped.
But this practice feels like a band-aid at best,
and yields even more ``leaning toothpicks.''

The above examples also illustrate how each level of embedding
can multiply the length of embedded strings by a factor of 2 or more,
yielding in the worst case
an exponential string-length explosion with the number of embedding levels.
While more than two levels of embedding may not be that common,
they do occur.

\xxx{
Convenience challenges.
Error-proneness issues.
Destructive interaction of inner and outer escaping mechanisms.

Shells, scripts, and command-line escaping.
Log files.

Larger languages.  JavaScriopt/TypeScript within HTML/XHTML.
Template-driven authoring languages (e.g., Hugo).
}

\subsection{A proliferation of quoting conventions}

Escaping issues such as those above have in part led many languages
to support multiple different types of quotes with different escaping rules.
Many languages allow string literals
to be either single-quoted or double-quoted,
so that quote characters of one type can be used literally
within string literals of the other,
as in \verb|"'"| or \verb|'"'|.

Some languages disable escape sequences in one form of string literal
so that backslashes may appear literally without multiplying in number:
\eg single-quoted Bourne shell strings (\verb|'\'| is the same as \verb|"\\"|)
or backtick-quoted raw string literals
in Go (\verb|`\`| is like \verb|"\\"|).
But without escapes it becomes more difficult to include
the forbidden terminating quote literally in the string:
typically one must compose multiple strings,
like \verb|"it's "+'"quoted"'|.
Some languages such as Python allow triple-quoted multiline strings
to make it less likely that the terminating sequence is needed in the literal:
\eg \verb|'''|$\dots$\verb|'''| or \verb|"""|$\dots$\verb|"""|.
But such a sequence may still need to appear, of course --
especially in written examples of exactly this syntax for example.

Some languages further mitigate this problem
by offering an effectively-unlimited number of delimiter pairs.
\href{https://docs.swift.org/swift-book/LanguageGuide/StringsAndCharacters.html#ID286}{Extended string literals in Swift}, for example,
surround a quoted sequence with a balanced number of \verb|#| signs:
\eg \verb|#"|$\dots$\verb|"#|, \verb|##"|$\dots$\verb|"##|, etc.
\href{https://www.lua.org/manual/5.1/manual.html}{Lua}
similarly offers \emph{long bracket} quotations
like \verb|[=[|$\dots$\verb|]=]|, \verb|[==[|$\dots$\verb|]==]|, etc.
This approach has the appeal that for any string to be delimited,
there always \emph{exists} some delimiter pair
that can quote it unambiguously.
But the delimiters must still be carefully matched to the quoted string,
or vice versa.
It is still not possible to embed \emph{any} string of a broad class
verbatim into \emph{any} ``hole'' or template in a host language
without thinking about, and potentially adapting,
either the choice of delimiters or the embedded string.
Further, the worst-case ``cost'' of embedding in terms of string expansion
still increases with each level of nesting --
in this case
at least only linearly, rather than exponentially, in the number of levels.
We would prefer, however, if embedding
required \emph{no} expansion with increased depth.

\xxx{ explore: PHP-generated HTML with inline JavaScript.
See for example \href{http://www.zedwood.com/article/how-to-properly-escape-inline-javascript}{How to properly escape inline javascript} }

\subsection{When security goes wrong}

The syntactic complexity of correctly embedding strings into code
has created several broad classes of security vulnerabilities,
where untrusted (typically user-entered) strings intended to be embedded
can maliciously ``trick'' an application into interpreting parts of the string
in the host language context.

SQL injection attacks~\cite{clarke12sql}, for example,
typically arise from the common practice
of embedding user-entered strings into string literals
within SQL query templates.
If a server composes an SQL query with a clause like
\verb|"WHERE name='"+userName+"'"|, 
and the untrusted \verb|userName| can be maliciously crafted
to contain an unescaped single quote,
then the attacker can prematurely terminate the SQL string literal
and add other SQL clauses, like \verb|OR '1'='1'|
to make the clause unconditionally true regardless of \verb|userName|.

Cross-site scripting (XSS) attacks~\cite{fogie07xss}
similarly exploit errors in the ubiquitous practice
of embedding content from an untrusted source --
such as fields from an HTML form --
into HTML or other markup without proper ``sanitization'' via escaping.
Suppose, for example, that one user of a Web-based discussion forum
can post a message containing an HTML \verb|<script>| tag --
which the discussion server then inserts into corresponding pages
viewed by \emph{other} users of the forum.
The injected JavaScript code can then potentially steal
authentication cookies or other private information
from \emph{all} users on the site
who might unwittingly read the maliciously-crafted message.

These and other broad classes of syntactic confusion attacks
have led to the security-critical practice
of sanitizing all potentially-untrusted user input
before embedding it into security-sensitive code of any kind --
whether SQL queries, HTML markup, or other host languages.
The forms of sanitization needed in a particular context
unfortunately tend to be complex and intricately dependent on
the syntax and semantics of the host language
that the untrusted content is to be embedded in.
Version updates in the host language or associated libraries,
which the application might not always track immediately,
can easily introduce new syntactic attack vectors
that the application
has not yet countered with appropriate sanitization logic.

We do not expect any syntactic discipline, including matchertext,
to eliminate the need to sanitize untrusted inputs.
If a future SQL query or Web form is designed to accept
embedded matchertext from an untrusted source, for example,
then it will likely still be security-critical to check
that the untrusted content \emph{is indeed valid matchertext},
and reject it if not.
However, a passive \emph{verification} like this
can be much simpler, and hence less bug-prone,
than a content-modifying \emph{transformation},
to escape all characters or sequences that might be ``sensitive''
in the host language.
Further, this security-critical check
could also be more uniform across host languages --
\ie checking only that the three ASCII matcher pairs are matched corrrectly
throughout the untrusted content,
rather than deeply verifying and/or transforming
based on the complex syntax of a particular host language.
Thus, while matchertext will not eliminate the need for sanitization,
it might tighten and simplify
the function of the most security-critical ``checkpoint'' --
namely checking that embedding content from an untrusted source
preserves the structural integrity
of the host language it is embedded in.

%% file: design.tex
\section{Matchertext design and rationale}
\label{sec:design}

This section first defines matchertext 
as an abstract mathematical syntax in \cref{sec:design:abstract},
then in \cref{sec:design:concrete}
as a concrete syntactic discipline
pragmatically inspired by predominant practices.
These definitions represent the ``core'' of the matchertext concept,
and are the \emph{only} rules
that different languages must ``agree on''
in order to achieve the main goal of escapeless interlanguage embedding.

\subsection{Abstract definition of matchertext}
\label{sec:design:abstract}

We first define \emph{abstract matchertext}
in order to ensure that the basic concept is clear and precise.

We assume at the outset we are given some arbitrary alphabet $\Sigma$,
along with some finite set $\Pi$ of character pairs 
$\{(o_1,c_1),\dots,(o_k,c_k)\}$,
such that $\{o_i, c_i\} \subseteq \Sigma$ for all $1 \le i \le k$.
We define the \emph{openers} $O$ as the set of characters $o$
such that some pair $(o, c) \in \Pi$.
Similarly, the \emph{closers} $C$ are the characters $c$
such that some pair $(o, c) \in \Pi$.
We assume and require that the sets of openers and closers do not overlap:
\ie $O \cap C = \emptyset$.
We define the \emph{matchers} $M$ as the set of all openers and closers
(\ie $M = O \cup C$).
We define the \emph{nonmatchers} $N$ as
all characters that are not matchers
(\ie $N = \Sigma \setminus M$)
A \emph{matchertext configuration}
is a pair $(\Sigma, \Pi)$ following the above rules.

We now define the language $L$ of \emph{matchertext strings}
inductively (\ie generatively) as follows:
\begin{itemize}
\item	Any string $n$ consisting exclusively of nonmatcher characters in $N$,
	including the empty string,
	is a matchertext string in $L$.
\item	For any matchertext strings $m_1,m_2,m_3$ in $L$,
	and for any pair $(o,c) \in \Pi$,
	the concatenation $m_1||o||m_2||c||m_3$
	is also a matchertext string in $L$.
\end{itemize}

Intuitively, this definition captures the basic rule
that \emph{matchers must match} in matchertext.
Openers can be introduced into valid matchertext
only when paired with a matching closer, and vice versa,
but nonmatchers may be interspersed throughout with no constraints.

We consider matchertext to constitute a purely syntactic rule
with no associated semantic meaning.
While we can formally define it as a syntactic language,
in practice we will refer to it as a \emph{syntactic discipline}
rather than a language
because it assigns no specific meaning or purpose to the strings in $L$,
and no structure apart from the basic rule that matchers must match.
The meanings and structural purposes of all characters --
both matchers and nonmatchers --
are deliberately left entirely open
for any particular ``matchertext-aware'' language to define.

\xxx{ formally prove
Always-embeddability property given that embedded string is valid matchertext.
}

\subsubsection{Escapeless embedding in matchertext}
\label{sec:design:abstract:embed}

Suppose there is a set of languages $\mathcal{L} = \{L_1,\dots,L_k\}$
whose members all agree on a common  matchertext configuration $(\Sigma,\Pi)$.
Any language $L_h \in \mathcal{L}$ can \emph{host}
embedded strings in any other language $L_e \in \mathcal{L}$
without escaping or other transformations to the strings from $L_e$,
provided $L_h$ enforces the following simple \emph{embedding rule}.
Any embedded matchertext string $m \in L_e$ 
must be delimited (surrounded) by
some pair of strings $s_o, s_c$ defined by the host language $L_h$,
such that the full embedding sequence is $s_o || m || s_c$ in $L_h$.
Further, $s_o$ must contain one or more open matchers in $O$
that would be unmatched in $s_o$ alone --
that is, $s_o$ alone \emph{cannot} be valid matchertext --
and $s_c$ must contain one or more corresponding close matchers in $C$
that would be unmatched in $s_c$ alone.

This embedding rule permits escapeless embedding
because no fixed characters or strings, including $s_o$ and $s_c$,
need be unconditionally forbidden in the embedded string $m \in L_e$,
provided that $m$ is valid matchertext.
The host language processor need not know anything
about the embedded language $L_e$ other than that $m$ is matchertext.
A host language processor that parses left-to-right from $s_o$
can ignore any embedded instances of $s_o$ and $s_c$ within $m$
because the matchers they contain must match within $m$.
The host language processor can unambiguously recognize
the closer string $s_c$ that terminates the embedding
because it contains at least one close matcher
that would be unmatched, and thus illegal, if it were part of $m$.

In order to guarantee that verbatim embedding always works reliably,
a host language not only can but \emph{must} refrain
from applying either transformations (\eg escapes)
or restrictions (\eg disallowing certain characters or sequences)
within the embedded matchertext it is hosting.
We can view embedded matchertext as analogous to a diplomatic embassy,
whose host country is required by international law to respect and protect
the ``involability'' of the embassy's ``premises''
and not enter or otherwise meddle in
the embassy's internal affairs~\cite[Article 22]{un61vienna}.
Beyond the matchertext rule that ASCII matchers must match,
an embedded language's syntactic affairs are exclusively its own,
not to be meddled in by a host language.
While exceptions to this rule may sometimes be justified
as we discuss below in \cref{sec:design:concrete:variations},
any exceptions inevitably reduce verbatim embedding compatibility.

\subsection{Concrete matchertext in practice}
\label{sec:design:concrete}

The abstract definition of matchertext above
and its basic structural rule apply in principle
to any matchertext configuration $(\Sigma,\Pi)$.
In practice, however,
we must standardize on particular choices of $\Sigma$ and $\Pi$
across a set of languages of interest
in order to achieve escapeless embedding among them.
We wish to identify a particular, concrete matchertext configuration
that fits existing syntactic syntactic practices as well as possible,
and facilitates escapeless embedding across 
minimally-adapted variants of today's popular machine-readable languages.

We therefore propose a \emph{standard matchertext configuration}
whose alphabet $\Sigma$ is the Unicode/UCS character set~\cite{iso10646ucs},
and whose matcher pairs $\Pi$ consist of
the ASCII parentheses \verb|()|,
square brackets \verb|[]|,
and curly braces \verb|{}|.

The choice of UCS as the character set $\Sigma$
is justified by the fact that machine-readable languages
have largely converged on this standard,
so it is in effect already decided.
In fact, the programming language community has also largely converged
on UTF-8 as the standard way to encode UCS plain text
into flat byte-stream source files --
although encoding is not a primary concern for matchertext
since it operates below the character set abstraction.

\subsubsection{Standardizing on matcher pairs}
\label{sec:design:concrete:standard}

A particular choice of the matcher pairs $\Pi$ is less obvious, however,
and hence demands more careful justification.
We start by ``deferring to authority'':
namely the authority embodied in
the UCS character set we already chose.
The parentheses, open brackets, and curly braces
are the only characters in the ASCII --
or ``\href{https://www.compart.com/en/unicode/block/U+0000}{Basic Latin}'' -- 
code block that are standardized as members of the
\href{https://www.compart.com/en/unicode/category/Ps}{Open Puntuation (Ps)}
and 
\href{https://www.compart.com/en/unicode/category/Pe}{Close Punctuation (Pe)}
character classes.
Exactly as their official names indicate,
these character classes denote characters whose standard purpose
is to serve as open and close punctuation in matched pairs.

\paragraph{Why not the ``angle brackets'' \texttt{<>}?}
Many programming language also use the ASCII characters \verb|<| and \verb|>|
in matching pairs,
such as for generic types in C++ and Java,
or markup in SGML-derived languages such as HTML and XML.
These characters are not standardized as open/close punctuation, however,
but as mathematical less-than and greater-than inequality symbols.
Further, they are used for this purpose in mathematical expressions,
in \emph{unmatched} fashion,
much more pervasively than their occasional use as matchers.
Requiring these characters to be matched in matchertext
would not only conflict with their primary standardized purposes
(\ie would ``defy the authority'' of UCS),
but would make it extremely cumbersome to express
standard mathematical inequalities
(\eg \verb|a < b| in \verb|if| expressions)
in almost all programming languages.
Omitting \verb|<>| from the matched pairs $\Pi$
of the standard matchertext configuration does not
conflict with or prevent their paired use in specific languages --
as we will see when we focus on SGML-derived languages later
in \cref{sec:host:ml} and \cref{sec:embed:ml}.
Omitting them from $\Pi$
means only that we do not impose a ``universal'' rule
that they \emph{must} be used \emph{strictly} as matchers,
without exception, throughout all valid matchertext.

\paragraph{Why only the ASCII open/close punctuation?}
The full UCS standard of course includes
much more open and close punctuation.
UCS also includes
\href{https://www.compart.com/en/unicode/category/Pi}{Initial Puntuation (Pi)}
and 
\href{https://www.compart.com/en/unicode/category/Pf}{Final Punctuation (Pf)}
character classes specifically for quotation marks
intended for use in pairs
(\eg «quote» or “quote”).
None of these extended UCS characters are commonly used
in machine-readable language syntax, however --
no doubt in part merely by tradition,
but also for the pragmatic reason that only the ASCII punctuation symbols
are directly typeable on most keyboard layouts.
Moreover, the open/close and initial/final punctuation
in the extended UCS blocks
do not occur strictly in pairs.
For example,
there are three different left double-quote characters
(codes 201C, 201E, and 201F)
that potentially match with the right double-quote character (code 201D),
depending on linguistic culture and typographical style.
Thus, deciding \emph{which} character pairs should or should not match
would become a much more complex question.
While specific languages are free to use any UCS punctuation
for their own language-specific structural or stylistic purposes,
it seems simplest and safest to restrict the matchertext set $\Pi$ of
\emph{strictly-matching} pairs to the ASCII matchers alone.

\paragraph{Why all three ASCII matcher pairs?}

We could of course be even more selective
in choosing the set of matcher pairs $\Pi$.
We could take only one matcher pair, for example:
either parentheses \emph{or} square brackets \emph{or} curly braces.
However, all of these matcher pairs are used quite pervasively,
in a variety of different structural roles in different languages,
and it is not readily apparent what principle would justify
choosing one of these matcher pairs over the others
to play a distinguished, globally-enforced matching role in matchertext.
Moreover, interspersing multiple distinct matcher pairs in structured text
in practice provides useful redundancy
that helps detect errors more quickly and localize them more precisely.
For example, it is much clearer where the missing close bracket is
in the string \verb|[{}([){}]|
than in the similar but more homogeneous string \verb|[[][[][]]|.
Finally,
any host language must use \emph{some} matcher pair
to delimit embedded matchertext strings,
as discussed above in \cref{sec:design:abstract:embed}.
Including all three ASCII matcher pairs in $\Pi$
thus gives languages maximum syntactic freedom
in defining the syntax of matchertext embeddings
(\ie a choice among three matcher pairs rather than a single prescribed pair).

\subsection{Matchertext configuration variations}
\label{sec:design:concrete:variations}

Even if adequately well-justified,
we cannot expect the standard matchertext configuration
as defined above to be a perfect or painless fit
for all situations in which string embedding is useful.
Including any matcher pair in $\Pi$ has the cost
of requiring those matchers to be escaped
in string literals and comments, for example,
as we detail later in \cref{sec:embed}.
There may be legitimate or even unavoidable reasons
to use other matchertext configurations in some cases,
keeping in mind that doing so reduces interlanguage embedding compatibility.

In general, deviations from the standard matchertext configuration
can be either \emph{tightening} (more restrictive),
\emph{loosening} (less restrictive), or a combination.

\subsubsection{Tightening variations}
\label{sec:design:concrete:variations:tight}

A matchertext configuration $(\Sigma',\Pi')$
is a \emph{tightening} of
the standard matchertext configuration $(\Sigma,\Pi)$ defined earlier
if it only removes characters from the alphabet ($\Sigma' \subseteq \Sigma$)
and/or makes additional matcher pairs sensitive ($\Pi' \supseteq \Pi$).
A string in the tightened matchertext configuration
may be copied verbatim to an embedding context
expecting the standard matchertext configuration,
but not necessarily in the other direction.

As we detail later in \cref{sec:host:uri},
uniform resource identifiers (URIs)~\cite{rfc3986} traditionally allow
only graphical characters -- and no spaces or control codes for example --
in order to make them manually transcribable.
To serve this transcribability purpose,
significant spaces and control codes must not appear \emph{anywhere} in a URI,
even in an embedded matchertext substring.
Thus, URIs may represent a justifiable use-case
for an alternate matchertext configuration that removes
the non-graphical characters from the alphabet.
This would unfortunately mean that a string cannot, in general,
be copied from a standard matchertext language into a matchertext URI
without transformation (\ie escaping spaces and control codes).

\subsubsection{Loosening variations}

A matchertext configuration $(\Sigma',\Pi')$
is a \emph{loosening} of
the standard matchertext configuration $(\Sigma,\Pi)$
if it only adds characters to the alphabet ($\Sigma' \supseteq \Sigma$)
and/or removes sensitive matcher pairs ($\Pi' \subseteq \Pi$).
A string in the standard matchertext configuration
may be copied verbatim to an embedding context
supporting the loosened matchertext configuration,
but not necessarily in the other direction.

A loosened matchertext configuration might be justified,
for example, if it is deemed critical to embed strings in some language $L_e$
that frequently makes unmatched uses of some ASCII matchers,
and the pain of escaping or otherwise adapting that syntax is deemed too great.

Mathematical notation, for example,
sometimes uses
``mismatched'' parentheses and square brackets
to represent half-open/half-closed intervals.
That is,
$[0,1)$ typically means any real number $r$ greater than or equal to zero
but strictly less than one ($0 \le r < 1$).
A machine-readable language making frequent use of this mathematical notation
might be considered too painful to embed
in a standard matchertext configuration,
and therefore might ``demand'' a looser configuration
in which perhaps only the curly braces are sensitive as matcher pairs.

This example seems fairly hypothetical, however,
as extremely few machine-readable languages
appear to support this mathematical
half-open/half-closed interval notation anyway.
Languages that do support some form of half-open/half-closed syntax
often do so with other, more matchertext-friendly notation.
Swift~\cite{apple22swift}, for example,
supports \href{https://docs.swift.org/swift-book/LanguageGuide/BasicOperators.html#ID73}{\emph{half-open range} syntax} like \verb|1..<4|
for the sequence of integers starting from and including 1,
up to but not including 4.
This syntax is perfectly compatible with the standard matchertext configuration
because it uses the mathematical inequality operators,
rather than unmatched matchers,
to express the range's open upper endpoint.

%% file: host.tex
\section{Host language considerations}
\label{sec:host}

This section focuses on considerations for,
and potential extensions to,
languages that may wish to \emph{host}
matchertext strings in other languages
and provide the convenience of ``cut-and-paste'' embedding.
\Cref{sec:embed} will later discuss considerations
for languages wishing to \emph{be embedded} conveniently.
Both sets of considerations are relevant
to languages wishing to be maximally ``matchertext-friendly'' of course.
We present host-language and embedded-language considerations separately,
however,
in order to emphasize their conceptual orthogonality:
a language could readily adopt hosting extensions but not embedding extensions,
or vice versa.

\subsection{General hosting considerations}

Suppose a host language $L_h$ wishes
to allow embedding a matchertext string $m$
from an arbitrary language $L_e$,
whose syntax is likely unknown to the host language.
In contexts where matchertext strings $m \in L_e$ are allowed,
$L_h$ must impose \emph{no} constraints on characters allowed in that context
other than the matchertext discipline (matchers must match).
Further, $L_h$ must not transform the embedded strings $m$ in any way
while extracting it from the host-language text.
Specifically, any escaping mechanisms or other transformations
that might normally apply to text in $L_h$
must be disabled in the context of the embedded string.
If any escaping mechanisms or other transformations
are active within the embedded string,
they must be those of $L_e$, not $L_h$.
We will see examples of this principle applied
in several specific contexts below.

Languages need not \emph{be} matchertext-compliant in their own syntax,
however,
just in order to \emph{host} embedded matchertext.
Existing languages can preserve full compatibility
with all their existing (non-embedded, non-matchertext) code --
continuing to allow unmatched matchers in string literals for example --
while incrementally adding extensions that make it easy
to embed matchertext strings verbatim within the host language.
This form of backward compatibility will likely be essential
to the incremental adoption of matchertext.

\begin{table*}
\begin{center}
\begin{footnotesize}
\begin{tabular}{l|l|l|l|c}
Class	& Description
	& Syntax
	& Example
	& See \\
\hline
	& & & & \\
C-like	& String escape
	& \verb|"|$\dots$\verb|\[|$m$\verb|]|$\dots$\verb|"|
	& \verb|"now \[quoting's "easy" in matchertext]"|
	& \ref{sec:host:c} \\
	& & & & \\
\ml	& Element
	& \verb|<|\textit{tag attrs}\verb| [|$m$\verb|]>|
	& \verb|<code [if (a<b) { printf("some <markup>\n"); }]>|
	& \ref{sec:host:ml:element} \\
	& Attribute
	& \verb|<|\textit{tag attr}\verb|=[|$m$\verb|]>|
	& \verb|<button onclick=[show("it's done!")]>OK</button>|
	& \ref{sec:host:ml:attr} \\
	& Section
	& \verb|<![MDATA[|$m$\verb|]]>|
	& \verb|<![MDATA[some example <b>bold</b> markup]]>|
	& \ref{sec:host:ml:section} \\
	& & & & \\
URI	& Bracket quote
	& \verb|[|$m$\verb|]|
	& \verb|http://trans.late/?page=[http://my.site/]&lang=en|
	& \ref{sec:host:uri} \\
	& Percent escape
	& \verb|%[|$m$\verb|]|
	& \verb|http://social.net/user/%[joe@email.net]/index.html|
	& \ref{sec:host:uri} \\
\end{tabular}
\end{footnotesize}
\end{center}
\caption{Summary of proposed matchertext hosting extensions.}
\label{tab:host:summary}
\end{table*}

We next examine languages with C-like string literal syntax
in \cref{sec:host:c},
then address SGML-derived languages such as HTML and XML
in \cref{sec:host:ml},
and finally in \cref{sec:host:uri}
we focus on uniform resource identifiers
in their role as a meta-syntax frequently ``hosting''
embedded identifiers derived from other syntaxes.

\Cref{tab:host:summary} summarizes the syntax extensions
for different language classes proposed in this section.
We emphasize that these are merely proposals for discussion.
Different language communities should and will make their own decisions,
and need not agree across languages on specific extension syntax
in order for matchertext to be useful.

\subsection{C-like host languages}
\label{sec:host:c}

An enormous variety of today's popular programming languages
are derived, either closely or loosely, from C~\cite{kernighan88c}.
Though differing widely in purpose, philosophy, and semantics,
a vast number of these C-inspired languages share similar syntax
for string literals.
In particular, most C-derived languages use
double and/or single quotes to delimit a string literal,
and backslash escape codes to insert ``special'' characters within the literal:
\eg \verb|"hello!\n"|.
Because quoted string literals
are the primary existing syntactic mechanism
for embedding (non-matchertext) strings traditionally,
they represent a natural starting point for considering matchertext extensions.

Given the ubiquity of backslash-escaped string literals,
we suggest that one reasonable extension for hosting matchertext
in C-like languages is via a new escape sequence,
such as \verb|\[|$m$\verb|]|,
where $m$ is arbitrary matchertext.
The embedded matchertext $m$ is uninterpreted by the host language processor
except to verify that ASCII matchers match
and to find the terminating close bracket.
Thus, quote characters, backslashes, whitespace, newlines,
or other control codes cease being ``special'' within the matchertext $m$ --
at least from the perspective of the host language.
For example, the string literal \verb|"\["'\]"|
becomes equivalent to \verb|"\"\'\\"|.
These and other characters might of course be ``special'' with respect to
whatever embedded language $m$ might be written in.

\subsubsection{Some syntactic alternatives}

The above proposal is only one of many possible alternatives of course,
which may be worth considering --
especially in the context of specific programming languages.
We now briefly discuss a few ``obvious'' alternatives
that seem less preferable for various pragmatic reasons.

The tradition of using quotes to delimit string literals
is unfortunate in terms of matchertext's ``cut-and-paste embedding'' goal.
Both the ASCII double quote (\verb|"|)
and the ASCII ``single quote'' (\verb|'|) --
technically standardized as an apostrophe and not a quote --
are ``undirected'' and do not come in matched pairs,
so C-style quoted strings do not naturally nest.
Unicode offers directed quote characters
intended for use in matched pairs,
but they are harder to type directly on most keyboards,
and are traditionally used in human-readable languages
rather than programming languages.
Also, the question of \emph{which} Unicode quote characters go together
is heavily language- and culture-dependent:
\eg “English”, „German“, «French», »Danish«, etc.
Thus, there is no obvious language-neutral way to choose and define
a particular set of Unicode directed-quote characters as matcher pairs.
Without doing that,
quote characters are not useful to host embedded matchertext,
because the ``matchers must match'' rule would not be sufficient
for the host language processor
to find the end of a matchertext string reliably.

Using ASCII matchers alone as new ``matchertext string literal'' delimiters --
like \verb|(|$m$\verb|)|, \verb|[|$m$\verb|]|, or \verb|{|$m$\verb|}| --
would obviously conflict with
many other long-established and doubtless higher-priority syntactic uses,
such as expression grouping \verb|a*(b+c)|, tuples \verb|(a,b)|,
lists \verb|[a,b]|, sets \verb|{a,b}|, and maps \verb|{a=1,b=2}|.

Nested \emph{combinations} of quotes and ASCII matchers --
such as \verb|["|$m$\verb|"]| or \verb|"[|$m$\verb|]"| or similar --
might also be initially appealing.
In most C-like languages, however,
such combinations would similarly conflict
with combinations of existing syntactic constructs
that are not unlikely to appear in existing code:
\eg a list whose sole element is a string literal, like \verb|["x"]|,
or a string literal containing brackets, like \verb|"[x]"|.
Using more deeply-nested matchers -- \eg \verb|[["|$m$\verb|"]]| --
only pushes these syntactic conflicts deeper
(a singleton list of a singleton list of a string literal).
Considering the other ASCII matchers (parentheses or curly braces)
does not improve the situation much.

Embedding matchertext in a string literal via a new escape sequence
also has the advantage that the \emph{entire} literal need not be matchertext.
Literals can mix matchertext with conventional literal text
including host-language escape sequences:
\eg \verb|"\t\[let's indent]\n\t\[a "quote"]"|.

The choice of square brackets for the proposed matchertext escape sequence
is somewhat arbitrary:
we could instead use use parentheses or curly braces,
or a longer sequence such as \verb|\m[|$m$\verb|]|.
Any choice may conflict with existing syntax in \emph{some} language:
\eg
\verb|\(|$m$\verb|)| conflicts with
\href{https://docs.swift.org/swift-book/LanguageGuide/StringsAndCharacters.html#ID292}{string interpolation in Swift},
\verb|\[|$m$\verb|]| conflicts with
\href{https://www.php.net/manual/en/language.types.string.php}{octal character escapes in PHP},
and
\verb|\{|$m$\verb|}| conflicts with
\href{https://ceylon-lang.org/documentation/1.3/reference/literal/string/}{Unicode escapes in Ceylon}.
Fortunately, different languages need not agree on
the precise syntax for hosting matchertext strings,
and can choose whatever syntax best suits that particular language.
To fulfill matchertext's main objective,
different languages need to agree \emph{only} on the basic rule
that the embedded string itself is arbitrary except that
ASCII matchers must match.

\subsection{SGML-style markup host languages}
\label{sec:host:ml}

While the venerable
Standard General Markup Languages (SGML)~\cite{iso8879sgml,goossens95sgml}
itself has waned in popularity,
its derivatives HTML~\cite{whatwg22html} and XML~\cite{w3c08xml}
are now ubuiquitous in Web content and programming.
Wherever the differences between these markup languages is not important,
we will refer to them all as \ml languages.

In their basic role as markup languages
used to produce rich, structured documents,
\ml languages frequently play ``host'' to embedded strings
in countless other languages:
typically, in the language(s) of software or APIs
that a marked-up document is written about.
Embedding code in other languages as verbatim text
is a basic and frequently-used purpose of HTML's
\verb|<code>| and \verb|<pre>| tags,
for example.
Beyond merely marking up verbatim text in other languages, however,
HTML in particular has evolved to include special-purpose support
for embedding several other languages within HTML:
namely scripting languages such as JavaScript or Tcl,
cascading style sheets (CSS)~\cite{w3c21css},
MathML~\cite{w3c14mathml},
and SVG~\cite{w3c18svg}.

The \ml languages are surprisingly complex syntactically,
especially given their simple-sounding purpose
of ``merely'' describing structured markup of usually human-readable text.
In particular,
there are at least three different syntactic contexts
in which strings in other languages are often embedded
into \ml languages --
and in which three different sets of quoting and escaping rules apply.
Embedded strings are often embedded
(1) as the content of an element,
(2) as an attribute within an element's start tag, or
(3) as verbatim text within a CDATA section.
We address each of these syntactic contexts in turn,
in each case suggesting potential matchertext extensions
that could help mitigate the various forms of ``escaping hell''
that these embedding contexts can create.

\begin{figure*}
\begin{center}
\begin{footnotesize}
\begin{tabular}{lrl}
\multicolumn{3}{l}{\textbf{(a) Embedding strings in other languages as element content, in standard HTML or with matchertext hosting extensions (+M):}} \\
& HTML	& \verb|<code>printf("Hello world!");/code>| \\
& +M	& \verb|<code [printf("Hello world!");]>| \\
& HTML	& \verb|<code>printf("Example &lt;b&gt;bold&lt;/b&gt; and &amp;bigstar; reference in HTML");]>| \\
& +M	& \verb|<code [printf("Example <b>bold</b> and &bigstar; reference in HTML");]>| \\
& HTML	& \verb|<script>document.getElementById("demo").innerHTML = "Hello world!";</script>| \\
& +M	& \verb|<script [document.getElementById("demo").innerHTML = "Hello world!";]>| \\
& HTML	& \verb|<script>document.getElementById("demo").innerHTML = "a <" + "/script> end tag";]>| \\
& +M	& \verb|<script [document.getElementById("demo").innerHTML = "a </script> end tag";]>| \\
\\
\multicolumn{3}{l}{\textbf{(b) Embedding strings in other languages within element attributes, in standard HTML or with matchertext extensions (+M):}} \\
& HTML	& \verb|<button onclick="okClicked()">OK</button>| \\
& +M	& \verb|<button onclick=[okClicked()]>OK</button>| \\
& HTML	& \verb|<button onclick="emitCharacter('\'')">Emit Apostrophe</button>| \\
& +M	& \verb|<button onclick=[emitCharacter("'")]>Emit Apostrophe</button>| \\
\\
\multicolumn{3}{l}{\textbf{(b) Embedding strings in within CDATA (character data) sections, in standard XHTML or with matchertext extensions (+M):}} \\
& XHTML	& \verb|<code>example <![CDATA[<b>bold</b>]]> markup</code>| \\
& +M	& \verb|<code>example <![MDATA[<b>bold</b>]]> markup</code>| \\
& XHTML	& \verb|<code>example <![CDATA[<![CDATA[character data]]]]><![CDATA[>]]> markup</code>| \\
& +M	& \verb|<code>example <![MDATA[<![CDATA[character data]]>]]> markup</code>| \\
& XHTML	& \verb|<code>example <![CDATA[<![CDATA[<![CDATA[double embedded]]]]]]>| \\
&	& \verb|<![CDATA[><![CDATA[>]]]]><![CDATA[>]]> markup</code>| \\
& +M	& \verb|<code>example <![MDATA[<![MDATA[<![MDATA[double embedded]]>]]>]]> markup</code>| \\
\end{tabular}
\end{footnotesize}
\end{center}
\caption{Examples of embedded strings in standard \ml languages,
	and with proposed matchertext extensions (+M).}
\label{fig:ml-emb}
\end{figure*}

\subsubsection{Strings embedded as element content}
\label{sec:host:ml:element}

One common form of embedding into \ml 
is marked-up text serving as the content of an element:
\eg example code between \verb|<code>| and \verb|</code>| tags
or between \verb|<pre>| and \verb|</pre>| tags in HTML.
Further, the \verb|<script>| and \verb|<style>| tags in HTML
exist specifically to embed scripting language code
and cascading style sheet (CSS) code, respectively,
as their content.

The syntactic rules governing
what can appear in text embedded as element content,
however,
depend intricately on the tag, the \ml language in question,
and even the language version.
In most elements such as \verb|<code>| and \verb|<pre>|,
any characters \verb|<| and \verb|&| appearing in the embedded string
must be escaped (as \verb|&lt;| and \verb|&amp;|),
to prevent the \ml parser misinterpreting them as
the start of a tag or a character reference,
respectively.
In XML, this rule applies to the content of all elements,
including the content of \verb|<script>| and \verb|<style>| tags
of XML-based XHTML.
In HTML, however, the content of \verb|<script>| and \verb|<style>| tags
is raw character data,
uninterpreted by the HTML parser except to find the end tag.
The content of such tags therefore \emph{can} contain
unescaped \verb|<| and \verb|&| characters --
and \emph{cannot} use HTML character entity references for escaping.
In HTML4, this uninterpreted content is terminated
by the first instance of a \verb|</| character sequence,
whether or not it is part of the corresponding end tag
(\verb|</script>| or \verb|</style>|).
HTML5 in contrast terminates the content with a sequence \verb|</|
followed by the appropriate end tag name.
In all of these cases, figuring out what \emph{must be},
what \emph{can be}, and what \emph{cannot be}
escaped is subtle and potentially confusing.

As a potential extension enabling any of the \ml languages
to host embedded matchertext conveniently,
we suggest the following new element syntax:

\begin{center}
\verb|<|\emph{name attributes }\verb|[|\emph{matchertext content}\verb|]>|
\end{center}

The \emph{name} and \emph{attributes} are the tag name and optional attributes
as they normally appear in a start tag,
and \emph{matchertext content} is the element content as literal matchertext
enclosed in square brackets,
uninterpreted except to find the end by matching matchers.
This syntax represents the entire element,
with no end tag,
so it is more concise than traditional start/end tag pairs.
Since the content within brackets is uninterpreted except to match matchers,
the content cannot contain further markup (child elements)
or \ml character entity references when using this syntax.

\Cref{fig:ml-emb}(a) illustrates a few examples
of embedding JavaScript into a \verb|<code>| or \verb|<script>| element,
either in standard HTML or with the proposed matchertext content syntax (+M).
The first and third examples embed trivial and non-problematic code.
The second example shows the embedding of literal HTML markup within HTML.
The fourth example illustrates the more troublesome corner case
where embedded JavaScript wishes to output
a \verb|</script>| end tag within a string literal.
Since HTML entity references are unavailable within a \verb|<script>| element,
the code must either use JavaScript escapes,
or construct the \verb|</script>| tag from two string literals,
to prevent the embedded literal from prematurely ending
the \verb|<script>| element.
In matchertext content syntax,
neither example is problematic and both are more concise.

\subsubsection{Strings embedded as attribute values}
\label{sec:host:ml:attr}

Besides element content,
scripting language code is often embedded in the attribute values
of \ml start tags,
most commonly to handle events in active user interface elements.
Attribute values represent a different syntactic context
in which different escaping rules apply.
When attribute values are delimited with single or double quotes,
the quote character that introduced the value must be escaped
(as \verb|&apos;| or \verb|quot;|)
if it is embedded in the attribute value.
Character references may appear and are substituted in attribute values,
like normal elements such as \verb|<code>| in HTML
but unlike \verb|<script>| or \verb|<style>| content.
As \href{https://www.w3.org/TR/html401/appendix/notes.html#notes-specifying-data}{the HTML specification notes},
this means that script and style data cannot be simply
cut-and-pasted between element content and attribute values
without care for the changed escaping rules.
HTML forgivingly allows \verb|<| and ``unambiguous'' \verb|&| characters
to appear unescaped in attribute values,
while XML requires them to be escaped (along with the active quote character).

One potential matchertext hosting extension
would be simply to allow square brackets as a third ``quoting style''
for attribute values,
where the text between the brackets is uninterpreted
except to match matchers and find the end.
With this extension as well as that above,
the quoting and escaping rules for matchertext element content
and matchertext attribute values would be identical,
allowing code to be cut-and-pasted between these contexts freely.

\Cref{fig:ml-emb}(b) illustrates
script text embedded in attribute values,
without and with matchertext hosting extensions.
The second example illustrates how any time
a string literal is needed in such embedded text,
the embedding effectively ``consumes'' both quote characters
in standard HTML or XHTML.
Matchertext embedding, in contrast,
preserves JavaScript's ``syntactic freedom''
of using one quote character to quote a verbatim instance of the other.

\subsubsection{Strings embedded in CDATA sections}
\label{sec:host:ml:section}

A third syntactic context in which strings are embedded in SGML and XML
(but not HTML)
is via CDATA sections of the form \verb|<![CDATA[|\emph{text}\verb|]]>|,
where \emph{text} is mostly-uninterpreted character data.
CDATA \emph{sections} are distinct from
CDATA-typed \emph{entities} or \emph{attributes}
as declared in an SGML document type definition (DTD)~\cite{english97cdata}.
CDATA sections offer the ``greatest protection''
from typical \ml escaping requirements,
in that \emph{only} the section-terminator sequence \verb|]]>|
is disallowed within the embedded text.
Because \ml escape sequences are unavailable within CDATA sections, however,
they also require the most-awkward syntactic contortions
in the hopefully-rare event that a \verb|]]>| sequence
needs to appear in an embedded string.
This ``worst-case scenario'' readily comes to pass
whenever one is \emph{writing about} CDATA sections and their issues
in a \ml markup language, for example.

A straightforward extension to host matchertext in a CDATA-like section
would be simply to add a matchertext section form
such as \verb|<![MDATA[|\emph{matchertext}\verb|]]>|,
where \emph{matchertext} is uninterpreted matchertext.
\Cref{fig:ml-emb}(b) illustrates three examples of markup
using CDATA sections versus corresponding MDATA sections.
The first example is simple and non-problematic in either case.
The second example illustrates how MDATA sections eliminate the problem
of embedding a \verb|]]>| sequence within such a verbatim section --
provided that matchers still match, of course.
The third example shows the more-extreme case of ``double embedding'' --
where the complexity and visual obfuscation of CDATA sections explodes,
while MDATA sections nest arbitrarily with no difficulty.
This double-embedding scenario might seem contrived,
but it is exactly what is needed, for example,
when attempting to write in \ml markup a visual example
(\eg in a \verb|<code>| block)
of the single-embedding problem and its typical ``preferred'' solution
of replacing \verb|]]>| sequences with \verb|]]]]><![CDATA[>| sequences
to ``close and reopen'' the outer CDATA section.

\xxx{ relevant:
\href{http://www.flightlab.com/~joe/sgml/cdata.html}{CDATA Confusion}
}

\subsection{Uniform Resource Identifiers}
\label{sec:host:uri}

Uniform resource identifier (URI) syntax~\cite{rfc3986}
has become a ubiquitous notation for naming and locating
not only web pages but innumerable other Internet resources.
As a ``small'' special-purpose syntax,
as opposed to a general-purpose programming language,
it is arguably more often useful as an embedded rather than a host syntax,
as we will focus on later in \cref{sec:embed:uri}.
In practice, however, innumerable other identifier syntaxes
get embedded into URIs regularly,
either as scheme-specific text
(\eg file names, phone numbers, cryptographic hashes),
or even as query parameter values.
This common and intentional use of URI syntax
as a uniform ``wrapper'' for other identifier syntaxes
makes URIs worth careful consideration as a potential host syntax
for matchertext embedding.

We suggest two syntactic extensions
to host matchertext within URIs,
which are potentially complementary
and could be adopted either together or individually.

\begin{figure*}
\begin{center}
\begin{footnotesize}
\begin{tabular}{lrl}
%\multicolumn{3}{l}{\textbf{URI examples without and with
%	matchertext percent-escape extension}} \\
%& URI	& \verb|http://social.net/user/joe%40email.net/index.html| \\
%& +M	& \verb|http://social.net/user/%[joe@email.net]/index.html| \\
& URI	& \verb|http://dev.site/myLibrary/doc/genericContainer%3CT%3E/api/| \\
& +M	& \verb|http://dev.site/myLibrary/doc/%[genericContainer<T>]/api/| \\
& URI	& \verb|http://search.engine/linksto?site=http%3A%2F%2Fmy.site%2F&results=50| \\
& +M	& \verb|http://search.engine/linksto?site=%[http://my.site/]&results=50| \\
& URI	& \verb|http://calculator.site/?expr=(1%2B2)*3%5E4%2F5| \\
& +M	& \verb|http://calculator.site/?expr=%[(1+2)*3^4/5]| \\
\end{tabular}
\end{footnotesize}
\end{center}
\caption{Examples of URIs without and with matchertext hosting extensions (+M).}
\label{fig:host:uri-examples}
\end{figure*}

First, extending the existing ``percent-encoding'' scheme
for escaping special characters
(\eg \verb|%20| to represent an ASCII space),
we suggest \emph{matchertext escape} sequences
of the form \verb|%[|$m$\verb|]|.
Like the backslash-escape form \verb|\[|$m$\verb|]|
suggested in \cref{sec:host:c} for C-like languages,
the matchertext $m$ is uninterpreted by the URI processor
other than to verify that matchers match
and find the terminating close bracket.
For example, \verb|file:///%[a<b>c`d]| becomes
a valid URI usable to access a local file named \verb|a<b>c`d|,
containing characters typically allowed in Unix-derived file systems
but traditionally forbidden in URIs.
Since percent-encoding by the host URI processor
is disabled within the embedded matchertext $m$,
\verb|%[100%]| becomes valid and equivalent to \verb|100%25|.
In effect, this matchertext escape syntax offers
a more concise, less obfuscated way to express arbitrary portions of URIs
in which several characters
would otherwise have to be individually percent-encoded.
\Cref{fig:host:uri-examples} shows a few examples of URIs
using conventional and matchertext percent-escapes for comparison.

Another potential syntactic extension is to allow
square-bracketed sequences \verb|[|$m$\verb|]|
to appear verbatim within the URI body,
where $m$ is otherwise-uninterpreted matchertext.
This is \emph{not} an escape sequence:
the square brackets are not eliminated in URI processing,
so \verb|[@]| is equivalent to \verb|%5B%40%5D|.
This extension essentially serves as a matchertext-friendly quoting syntax
that may be used in specific URI schemes,
or within pathname components or query strings,
to embed substrings in other identifier syntaxes (even other URIs)
without obfuscation.
We will explore the usefulness of this extension further
when we consider \emph{matchertext resource identifier} or MRI syntax
later in \cref{sec:embed:mri}.
This extension is backwards-compatible with existing (valid) URIs
because current syntax permits brackets \emph{only}
in special-purpose IPv6 address syntax as part of the authority field.

\xxx{discuss issue of spaces? maybe later with MRIs}

\subsubsection{Precedents for URI syntax extensions}

Originally standardized as uniform resource locators or URLs~\cite{rfc1738},
URIs traditionally allow only a small subset of ASCII characters
to appear verbatim.
Non-graphical characters,
or those deemed ``unsafe'' for various reasons,
must be escaped via percent-encoding.
The set of allowed characters, and their purposes,
has been ``liberalized'' multiple times historically, however.

IP version 6 introduced colon-separated hexadecimal addresses
(\eg \verb|1234::abcd|~\cite{rfc2373}),
which conflicted with the URI's use of the colon to separate
an IP address from a port number (\eg \verb|http://1.2.3.4:80/|).
The square brackets \verb|[]| were therefore shifted
from forbidden to ``reserved'' characters in URIs,
for use \emph{only} in embedding IPv6 addresses
into the ``authority'' field of URIs,
like \verb|http://[1234::abcd]:80/|~\cite{rfc2732}.

International Resource Identifiers or IRIs~\cite{rfc3987}
further liberalized URI syntax,
allow most of the graphical characters
in the extended Unicode/UCS character sets to appear verbatim in URIs.
IRIs preserved backwards compatibility
in part by defining standard conversion processes back and forth
between internationalized IRIs and legacy ASCII-only URIs.
The set of ASCII characters allowed in IRIs remained tightly restricted,
however.

Still later, IPv6 introduced scoped identifier syntax,
allowing an interface number or name to be specified with an IPv6 address,
\eg \verb|1234::abcd%1| or \verb|1234::abcd%if0|~\cite{rfc4007}.
This new syntax again conflicted with URI syntax,
leading to further syntactic hacks involving
mandatory percent-escaping of the percent sign
indicating a scoped identifier~\cite{rfc6874}.

These points in in URI evolution illustrate
a repeating precedent for liberalizing URI syntax
to accept previously-forbidden characters
and to make URIs more ``friendly'' and accommodating
of embedded strings derived from other languages --
whether machine-readable (\eg IPv6 addresses)
or human-readable (international languages).
Adopting matchertext hosting extensions such as those above,
permitting URIs to host other syntaxes more cleanly
without the traditional  syntactic hacks and percent-encoding obfuscation,
could be a useful step in allowing URIs to fulfill their ambition
of being a uniform ``meta-syntax'' framework
accommodating an unlimited variety of specific identifier syntaxes.

\xxx{ There are probably many, many other examples
in RFCs and elsewhere on ugly syntactic hacks to embed
random syntax X into URIs.  Make a list and summarize it briefly.}

\subsubsection{How liberal to liberalize?}
\label{sec:host:uri:liberal}

The above considerations, however,
raise the obvious question:
\emph{how far should liberalization of URI syntax go?}
Beyond the square brackets,
which \emph{other} characters that were previously disallowed in URIs and IRIs
eventually be permitted, and in what contexts?

In terms of our current focus on hosting embedded matchertext,
the ideal would clearly be to allow \emph{any} UCS characters in URIs --
at least within a matchertext escape \verb|%[|$m$\verb|]|
or a matchertext quote \verb|[|$m$\verb|]|.
This approach would clearly provide the maximum latitude
for embedding other syntaxes into URIs cleanly in the future.
Further, only this ``extreme liberalization''
would guarantee that \emph{any} matchertext,
from any language conforming to the standard matchertext configuration
(\cref{sec:design:concrete}),
may be embedded verbatim into a URI without escaping.

As briefly discussed earlier in \cref{sec:design:concrete:variations},
however,
this arguably might be ``going too far'' in the case of URIs.
We first consider the graphical ASCII characters
that are currently disallowed in URIs,
then the non-graphical characters such as spaces and control codes.

\paragraph{Graphical characters:}

URIs traditionally forbid
``angle brackets'' \verb|<>| and double quotes \verb|"|
from use within URIs,
because these characters are sometimes
used to delimit URIs in surrounding freeform text:
\eg \verb|<http://my.site/>| or \verb|"http://my.site/"|.
In any ``legacy'' URI parsing context unaware of matchertext extensions,
the appearance of these characters within a URI might indeed
prematurely terminate the recognized URI,
an issue we will return to later in \cref{sec:embed:uri:end}.
In a context aware of the matchertext extensions, however,
there is no syntactic ambiguity between angle-brackets or double-quotes
used to surround a whole URI
and any that may appear within embedded matchertext.
The matchers delimiting the embedded matchertext
unambiguously serve to differentiate "inside" from "outside":
\eg as in \verb|<http://my.site/%[>"<]>|.

The original URL standard in 1994~\cite{rfc1738}
additionally declared all the following characters 
to be ``unsafe'' in URIs 
``because gateways and other transport agents are known to sometimes modify
such characters'':

\begin{center}
\verb|{ } |\texttt{|}\verb| \ ^ ~ [ ] `|
\end{center}

Nearly 30 years later,
the ``gateways and other transport agents'' that text containing URIs
tend to pass through have no doubt evolved drastically
(or been replaced entirely).
It is far from clear, therefore,
that URIs today face the same modification perils as those of 1994.
The text justifying the exclusion of these characters
was dropped from the latest URI standard~\cite{rfc3986},
though the characters themselves remain forbidden
apart from the square brackets.
IRIs introduced thousands of other new characters into the allowed set
without dire consequences.
With adequate care taken for backwards compatibility (as was done with IRIs),
it may be high time to consider allowing
the rest of the ASCII printing characters above into URIs --
\emph{at least} in embedded matchertext hosted within URIs.

\paragraph{Non-graphical characters:}

URIs also traditionally disallow non-graphical characters such as spaces,
as well as control characters such as newlines and tabs,
for a different purpose: the goal of \emph{transcribability}.
It was difficult in 1994 to transcribe by hand
a string containing spaces, newlines, tabs, or other control codes,
and it is probably just as difficult to do so today.
In an era of proliferating QR codes,
the only question might be to what extent manual transcribability
is still a crucial goal for URIs.

Nevertheless,
manual transcription remains an important and not-uncommon use of URIs,
and compromising the transcribability goal
would arguably represent a much more fundamental shift
in the principles underlying URIs
than decisions about allowing or disallowing particular printing characters.
This consideration therefore suggests that we
cease the liberalization of URIs just short of
allowing non-graphical characters.

Further, the transcribability goal is served
only if the \emph{entire} URI is readily transcribable,
including any embedded matchertext substrings it may contain.
Thus,
as briefly mentioned earlier in \cref{sec:design:concrete:variations:tight},
the URI context may justify a \emph{tightened} matchertext configuration
whose alphabet $\Sigma$ is reduced to exclude non-graphical characters.
The cost is that arbitrary matchertext is not necessarily copyable verbatim
from C-like languages into URIs,
but this compatibility cost may be justified in this case.

\subsection{Regular Expression Syntax}
\label{sec:host:regex}

Despite being a ``small'' synax for the special purpose
of matching patterns in strings,
regular expressions (REs) and their use are complex enough in practice
that multiple entire books have been written
about them~\cite{fitzgerald12introducing,friedl06mastering,goyvaerts12regular}.
Two regular expression syntaxes were standardized
by POSIX~\cite{ieee17posix},
while the \href{https://www.perl.org}{Perl language} and the
\href{https://www.pcre.org}{Perl Compatible Regular Expressions (PCRE) library}
it inspired
define advanced syntax that has become popular
in numerous other languages and applications.

In filling their basic role of matching patterns in text,
REs must inherently must embedd strings
comprising the patterns to be matched --
and those embedded strings can be in any syntax.
Thus, being able to host embedded strings of any language
with minimal obfuscation
in principle facilitates an RE's basic pattern-matching role.

Since the most common RE syntax uses backslash escapes
similar to those in C-like languages,
the same matchertext escape extensions,
such as \verb|\[|$m$\verb|]|, could work for REs.
Because of the large number of punctuation characters
that are sensitive in REs, however,
some popular RE syntaxes such as
\href{https://www.pcre.org/original/doc/html/pcrepattern.html}{PCRE}
guarantee the rule that if a backslash
``is followed by a character that is not a number or a letter,
it takes away any special meaning that character may have.''
This way, a user can just conservatively
escape \emph{all} literal punctuation appearing in an RE, 
instead of remembering which punctuation \emph{must} be escaped.
Preserving this rule may suggest a longer, letter-based
matchertext escape sequence
such as \verb|\m[|$m$\verb|]|.

Another alternative would be to use some non-backslash-escape syntax,
such as \verb|{{|$m$\verb|}}|.
This alternative syntax uses the same curly braces
that REs already commonly use
for repetition quantifiers like $c$\verb|{1,3}|,
but in a syntactically non-conflicting fashion
since the inner braces cannot be mistaken for repetition quantifiers.
With this alternative syntax,
the pathological ``\href{https://en.wikipedia.org/wiki/Leaning_toothpick_syndrome}{leaning toothpick syndrome}'' example RE,
matching the double-backslash \verb|\\| in a UNC name (see \cref{sec:bg}),
becomes the more readable RE \verb|{{\\}}|.
Embedding this RE in a C-like string literal with matchertext extensions
in turn becomes \verb|"\[{{\\}}]"|,
for a more manageable three leaning toothpicks total,
in contrast with the traditionally-required eight (\verb|"\\\\\\\\"|).

%% file: embed.tex
\section{Embedded syntax considerations}
\label{sec:embed}

Having focused above on host language considerations,
we now switch focus to considerations for languages
to \emph{be embedded} as matchertext.
The languages of interest for embedding
overlaps heavily with those of interest as host languages;
we separate these discussions mainly to emphasize the orthogonality
of host- and embedded-language issues and cleanly separate them.

It is already readily feasible to write valid matchertext
in most of the languages we will consider for embedding.
This is because most popular machine-readable languages
already largely conform to the ``matchers must match'' rule
in their explicit uses of the matcher characters.
Violations of the matchertext rule most commonly occur
only in embedded ``free-form'' text such as string literals and comments.
The language extensions we will propose are motivated almost exclusively
by increasing convenience and visual clarity,
and are by no means essential.

\subsection{String literals in C-like languages}
\label{sec:embed:c}

Almost certainly the most common context in which unmatched matchers
appear in most today's existing source code is within string literals.
This is especially true of code to print, or parse,
machine-readable code in almost any syntax.
Structured pretty-printing code frequently includes code sequences like this:

\begin{quote}
\verb|print("[")| \\
\emph{output all elements of a list} \\
\verb|print("]")|
\end{quote}

Similarly, parsing code often uses \verb|if|, \verb|switch|,
or \verb|case| conditionals
to recognize and parse matcher-delimited syntactic structures,
as in:

\begin{quote}
\verb|if peekNextChar() == '[':| \\
\verb|  scanChar('[')| \\
\verb|  |\emph{scan all elements of a list} \\
\verb|  scanChar(']')| 
\end{quote}

Printing and scanning code like this generally violates the matchertext rule,
and adapting such code most likely represents the biggest ``pain point''
in any venture to write readily-embeddable matchertext.

Almost all programming languages already offer a workable
if slightly cumbersome solution:
simply replace unmatched matchers in string literals
with suitable numeric character escapes.
Instead of \verb|print("[")|, for example,
write \verb|print("\x5B")| (C, C++, JavaScript)
or \verb|print("\u005B")| (Java, JavaScript, Go).
This always works;
the main annoyance is that it requires the writer (and reader) of the code
to remember or look up the codes for the matcher characters in an ASCII table.

The usual solution in C-like languages
to handle ``special'' characters in string literals
is simply to backslash-escape the special character,
like \verb|\[|.
This traditional solution does not work for unmatched matchers in matchertext,
however,
because the matchertext rule is deliberately language-independent
and oblivious to language-specific syntax such as that of string literals.
So a backslash-escaped unmatched bracket \verb|\[|
remains just as much a matchertext violation as the bracket alone.

There is a solution that avoids the need for ASCII tables, however.
Because literal matchers are a problem in matchertext only when unmatched,
we can simply introduce escape sequences that incorporate
\emph{both} matchers as a properly-matched pair,
while ``selecting'' only the opener or closer of the pair.
In C-like languages, for example,
we suggest the sequence \verb|\o()| to escape an open parenthesis,
\verb|\c()| to escape a close parenthesis.
Similarly,
\verb|\o[]| and \verb|\c[]| represent open/close square brackets,
and \verb|\o{}| and \verb|\c{}| represent open/close curly braces.

The choice of the letters \verb|o| and \verb|c| to escape the matchers
is consistent with their standardized character classes:
\href{https://www.compart.com/en/unicode/category/Ps}{``Open Puntuation (Ps)''}
and
\href{https://www.compart.com/en/unicode/category/Pe}{``Close Punctuation (Pe)''},
respectively.
We might consider \verb|l| and \verb|r| for ``left'' and ``right'',
escept \verb|\r| is a near-universal escape for carriage return (CR).
A few languages already use \verb|o| or \verb|c| in escape sequences:
\eg Raku uses \verb|\o[|$n$\verb|]|
to denote the ASCII character with octal value $n$,
and uses \verb|\c[|$n$\verb|]|
to denote a Unicode character with name or decimal value $n$.
Many of these existing uses are technically not in conflict syntactically,
provided the existing use requires a non-empty string between the matchers --
as Raku does in the above cases, for example.
% see raku/escapes.raku for a trivial script with which to verify this.
In any case, different languages need not agree
on specific escapes sequences for unmatched matchers
and are free to make their own stylistic choices.

\xxx{relationship: triple-quoted/multiline literals}

\subsection{Comments and derived documentation}

Another context in which unmatched matchers may regularly appear
in typical source code is within comments:
\eg as part of human-readable text \emph{describing}
how the associated code handles particular characters.
Conventional language processors usually just ignore unmatched matchers
(along with everything else) in a comment.
But the matchertext discipline operates below and oblivious to
the syntax of a particular language,
and hence does not know what a ``comment'' is --
so the matchertext discipline must disallow unmatched matchers even in comments.

Since comments are generally intended for humans reading the source code,
it is usually possible simply to rephrase the comment
to avoid a literal use of unmatched matcher characters:
\eg just name it (`open parenthesis')
instead of writing it (`\verb|(|').
Another alternative,
if a language adopts the above extensions for string literals,
is simply to use these matchertext-friendly escapes in comments as well
(\eg \verb|\o()|).

In some languages,
comments often get used to produce API documentation,
using tools like \href{https://www.oracle.com/java/technologies/javase/javadoc-tool.html}{Javadoc}
or \href{https://pkg.go.dev/golang.org/x/tools/cmd/godoc}{godoc}.
In such cases,
it may be useful to interpret escape sequences such as those above
while auto-generating documentation from source code,
so that a documentation comment like `\verb|// Parse a \o()|'
becomes `Parse a (' in the formatted output generated from the code.

\subsection{SGML-derived languages}
\label{sec:embed:ml}

Considerations similar to those above for string literals
apply when we wish to embed \ml-language markup
into other languages as matchertext.
The most common reason unmatched matchers appear in markup
is when needed in literal text being marked up:
\eg human-readable text \emph{about} the matcher characters
or syntactic constructs built from them,
or code examples that contain unmatched matchers.

As with C-style string literals,
\ml languages already offer a workaround:
simply use character references,
either named (like \verb|&lpar;|)
or numeric (\verb|&#x0028;|).
For the same reasons as above,
we may like to have extensions
offering more visually-obvious alternatives for writing matchertext:
\eg \verb|&o();| and \verb|&c();|
for open and close parentheses, 
respectively.

\subsection{Uniform resource identifiers}
\label{sec:embed:uri}

Since uniform resource identifier (URI) syntax represents
a special-purpose ``little language'' just for expressing identifiers,
URIs are predominately embedded in other contexts --
software source code, documentation markup, configuration files, etc.
Especially since URIs are intended to be human-readable,
it would thus seems useful if URIs
could be maximally ``friendly'' for embedding.

\subsubsection{The near-matchertext-compliance of URIs}

Conventional URI syntax~\cite{rfc3986}
already ``nearly'' complies with the ``matchers must match'' rule
and is thus, usually, embeddable verbatim in a matchertext context.
Curly braces are formally disallowed in URIs.
Square brackets are allowed \emph{only} to surround IPv6 addresses
in the authority field,
in properly-matched fashion.
Thus, the only unmatched matchers that \emph{can} exist
in a strictly-valid URI are parentheses.
Even these, when appearing in URIs,
often still come in matched pairs anyway.\footnote{\begin{tiny}
	For example:
	\texttt{https://en.wikipedia.org/wiki/URI\_(disambiguation)}
	\end{tiny}}

In the rare cases when unmatched parentheses are ``needed'' in a URI,
they may always be percent-escaped as \verb|%28| or \verb|%29|.
For example, the string `\verb|open(|'
becomes `\verb|open%28|' in a matchertext URI,
`\verb|close)|'
becomes `\verb|close%29|',
and `\verb|close)open(|' becomes
`\verb|close%29open%28|'.
The string `\verb|open(close)|'
need not be rewritten at all in a matchertext URI,
since the matchers it contains already happen to match.

We could always consider escaping extensions
such as \verb|%o()| and \verb|%c()|,
but it is far from clear that their likely-marginal need
would justify the syntactic complexity in this case.
Even if URI syntax is liberalized further to allow
square brackets and/or curly braces in components,
it is unclear how commonly unmatched matchers would be needed,
since it is not particularly common to write parsing or scanning code
within a URI for example.

\subsubsection{The URI end-finding problem}
\label{sec:embed:uri:end}

Nevertheless, 
URI syntax does suffer from at least one significant usability flaw
arising from its frequent use as an embedded syntax.
URIs can and often do appear almost ``anywhere''
in freeform human-readable text --
\eg typed or copied into E-mails, notes, documents, etc.
Smart text editors often try to detect URIs on entry
and automatically turn them into hyperlinks --
but these heuristics can easily break because
there is no unambiguous syntactic separation between the URI
from surrounding (particularly following) text.
Suppose for example that I type or copy this text into an E-mail:

\begin{footnotesize}
\begin{quote}
\verb|My site is https://bford.info/index.html.|
\end{quote}
\end{footnotesize}

The trailing period (\verb|.|) \emph{could} be part of the URI,
but in this case was probably intended to terminate my English sentence.
I could try to ``armor'' the URI, like this:

\begin{footnotesize}
\begin{quote}
\verb|See my site (https://bford.info/index.html).|
\end{quote}
\end{footnotesize}

But the close parenthesis, as well, \emph{could} be part of the URI
and be sucked into the link by a ``greedy'' URI auto-recognizer,
resulting in a broken link.
A careful reader of Appendix C of the URI specification~\cite{rfc3986}
might find the recommendation to delimit URIs
with angle brackets \verb|<>| --
but rather few people seem to be aware of this recommendation in practice,
let alone are following it.

\subsubsection{Matchertext resource identifiers (MRIs)}
\label{sec:embed:mri}

Given how commonly URIs are embedded in both freeform human-readable text
as well as other machine-readable syntaxes of all kinds,
we suggest that a more useful and ambitious potential evolution
would make URI syntax \emph{self-delimiting}.
In particular,
let us consider an alternative potential URI syntax
in which we surround the URI's body -- everything after the scheme name --
with square brackets instead of separating it from the body with a colon.
Thus, \verb|http://my.site/| becomes \verb|http[//my.site/]|.
This alternate syntax uses only characters
that are already used (and reserved) in current URI syntax,
and it remains readily recognizable in freeform embedded contexts,
but now the end can always be found unambiguously with no heuristic guessing.

Let's call this new syntax
a \emph{matchertext resource identifier} or MRI.
Since MRI syntax is distinct and not readily confused with traditional URIs,
it could enforce the rule that all URI body content within the brackets
must be matchertext --
\ie that unmatched matchers in the body must be percent-encoded --
for verbatim embedding of other syntaxes (or other MRIs) in the body.
Just as IRIs~\cite{rfc3987} liberalized URI syntax
while preserving backward compatibility
by defining automatic conversions in both directions,
MRI syntax could similarly be converted automatically
to or from traditional URI and IRI syntax.

Assume that MRI syntax includes the extensions
discussed earlier in \cref{sec:host:uri} --
in particular the rule that a square bracket sequence \verb|[|$m$\verb|]|
nested within the URI body protects the embedded matchertext $m$
from percent-encoding in the outer context.
With this syntax, MRIs cleanly nest with no escaping needed,
not even to introduce a matchertext embedding context.
An embedded MRI appearing in a path or query string component
of a host MRI never need be escaped, for example,
as illustrated by the examples in \cref{fig:search-query}.

Moreover, MRI syntax could potentially be \emph{simpler}
than traditional URI syntax,
because complex and rarely-used sub-syntaxes such as IPv4 and IPv6 addresses
could be ``broken out'' of the main MRI syntax
and handled instead as embedded MRIs in the host MRI's authority field.
For example,
the URI `\verb|http://1.2.3.4:80/|' would become
the 2-level MRI `\verb|http[//ip4[1.2.3.4]:80/|', and
the URI `\verb|http://[1234::abcd]:80/|' would become
the MRI `\verb|http://ip6[1234::abcd]:80/|'.
%IPv6 addresses with scoped identifiers
%could avoid obfuscation in URIs~\cite{rfc6874}:
%the URI `\verb|http://[1234::abcd%25eth0]/' becomes
%the MRI `\verb|http://ip6[1234::abcd%eth0]/'.
The MRI host field syntax thus knows only about domain names or nested MRIs,
and not about IP address syntax.

\subsection{Regular expressions}
\label{sec:embed:re}

Typical regular expression (RE) syntax is C-like
in terms of using backslashes to escape sensitive punctuation characters
within text to be matched.
Similar escape sequences like \verb|\o| and \verb|c| for unmatched matchers
could therefore be introduced as discussed above in \cref{sec:embed:c}.

One complication is that the popular
\href{https://www.pcre.org/original/doc/html/pcrepattern.html}{PCRE syntax}
already uses \verb|\o{|$n$\verb|}|
for character escapes with octal numeric value $n$.
This octal-escape usage of \verb|\o|
technically does not conflict syntactically
with \verb|\o{}|, however,
since the $n$ in an octal escape cannot be the empty string.

PCRE syntax also offers \verb|\c|$x$ as a way to enter control characters,
by flipping bit 6 of ASCII character $x$.
This is a syntactic conflict with the proposed matchertext escapes,
but perhaps a tolerable one.
The sequence \verb|\c(| would constitue
a bizarre and unlikely way to express a simple literal letter `\verb|h|',
and \verb|\c{| would be a strange synonym for a semicolon `\verb|;|' --
neither of which need escaping at all.
The sequence \verb|\c[| might be slightly more likely to see use
to express the ASCII escape (ESC) control code (hex 1B) --
but PCRE already provides the more-concise and obvious sequence \verb|\e|
to express this control code.

\subsubsection{Character classes}
\label{sec:embed:re:class}

Backslash escapes are normally disabled in
bracketed RE character class notation like \verb|[a-z0-9]|.
The matchertext discipline does not present a problem
when expressing a character class containing \emph{both} matchers of a pair.
For example,
the character class \verb|[()[]{}]| matches any matcher character,
while \verb|[^[]]| matches anything but a square bracket.
Including just one unmatched matcher in a character class
becomes less convenient, however.
A slightly-cumbersome workaround
is simply to shift unmatched matchers outside the character class:
\eg \verb|[a-z{]| might be rewritten as \verb=([a-z]|\o{})=.

A more-appealing syntactic extension might be to introduce the rule
that a less-chan character \verb|<| in a character class,
when immediately surrounded by a pair of matchers,
``selects'' only the open matcher for literal inclusion.
The example above therefore becomes \verb|[a-z{<}]|.
A greater-than character \verb|>| between a matcher pair
similarly selects only the close matcher:
\verb|[^[>]]| matches anything but a close bracket.
One might view the \verb|<| or \verb|>| character either
as a matchertext-insensitive angle-bracket
``standing in'' for the desired sensitive matcher,
or as an arrow ``pointing'' left or right to the desired matcher.

\xxx{
Regular expressions in
XXX %\href{https://www.tcl.tk/man/tcl8.4/TclCmd/re_syntax.html#M32}{Tcl}
and
XXX %\href{https://developer.mozilla.org/en-US/docs/Web/JavaScript/Guide/Regular_Expressions/Character_Classes#types}{JavaScript}
also use \texttt{\\c}$X$ as an escape sequence for control code characters,
though with different precise rules.
Using this syntax with an opener as the character $X$
appears to be illegal in JavaScript regular expressions,
and perhaps legal but unlikely to be used in Tcl regular expressions.
}

\xxx{
Question: how often are these escape sequences used,
and how often are the equivalent sequences for introducing control codes used
(\eg single-letter sequences, or octal or hex numeric sequences)?
Which specific escape sequences does this escape sequence get used for
and with which characters $X$?

Relevant: \href{https://www.tcl.tk/man/tcl8.5/tutorial/Tcl21.html}{More Quoting Hell - Regular Expressions 102}.
Also \href{https://wiki.tcl-lang.org/page/Quoting+hell}{Quoting hell}
}

\xxx{Trouble spots in syntax tradition.
When are unmatched matchers traditionally used
in actual language syntax, not just as literal text embedded within syntax?
Mathematical half-open range/set notation.
How problematic is this?
Other examples?
}

\xxx{Future: other oft-embedded languages to look at.  For example:
	SQL
	JSON
}

%% file: impl.tex
\section{Implementations of matchertext}
\label{sec:impl}

This section is mainly a placeholder at the present.
Some preliminary work is underway
implementing experimental matchertext extensions
for several languages and embedding-oriented syntaxes.
This section is intended to be expanded
as we gain experience
implementing and using matchertext extensions.

For those wishing to help with implementation and experimentation,
the following are some of the key work items
enabling us to start experimenting with matchertext in the context
of any particular language of interest:
\begin{itemize}
\item	A brief specification of proposed
	syntactic extensions for hosting matchertext
	and/or conveniently writing embedded matchertext,
	adapted to the specific language of interest,
	with clearly-defined rationale for the particular syntax choices.
\item	An implementation of those extensions for hosting and/or embedding,
	selectively enabled via configuration parameters
	for backward compatibility,
	in some mature processor for the language of interest
	(\eg a compiler, interpreter, or library implementation).
\item	A configurable extension to the language processor
	that optionally causes it
	to check and enforce the matchertext discipline
	in processed text:
	\ie to verify that matchers match in source files or strings.
\item	Purely for experimentation purposes,
	an extension of some processor for the language
	that can analyze ``legacy'' source files in the language
	to detect and categorize matchertext violations
	(\eg whether in string literals, comments, or elsewhere).
	We will use this to help estimate the likely ``pain''
	of adopting the matchertext discipline in the language
	and how commonly this pain would affect typical code today.
\end{itemize}

\subsection{MinML: minified matchertext markup}

One early experiment in matchertext-friendly syntax design
is MinML~\cite{ford22minml},
an alternative syntax for SGML-derived markup languages like HTML and XML.

Beyond merely adding matchertext hosting and embedding extensions
as discussed in the sections above,
MinML more ambitiously reformulates the basic \ml syntax
to rely on matching brackets for basic structure
rather than matching start/end tags as in SGML tradition.
For example, \emph{emphasis} is written like
\verb|em[emphasis]| rather than \verb|<em>emphasis</em>|.
Character references are written like
\verb|[star]| instead of \verb|&star;|.

A ``quotation'' delimited by matching quote characters
may be written like
\verb|"[quotation]| in MinML instead of \verb|&ldquo;quotation&rdquo;|.
A comment is \verb|-[comment]| instead of \verb|<!--comment-->|.
A raw embedded text sequence
is written \verb|+[verbatim]| instead of \verb|<![CDATA[verbatim]]>|.
MinML's embedded sequences leverage matchertext
to support arbitrary nesting,
so a verbatim example of a raw matchertext sequence
is simply \verb|+[+[example]]|, rather than in XML:

\begin{footnotesize}
\begin{center}
\verb|<![CDATA[<![CDATA[example]]]]><![CDATA[>]]>|
\end{center}
\end{footnotesize}

For escaping unmatched matchers,
MinML supports both the traditional HTML named and numeric character references,
and bracket-delimited versions of the ``visual'' matcher escapes
suggested earlier in \cref{tab:unmatched-matchers}
and \cref{sec:embed:re:class} discussing regular expressions:

\begin{center}
\begin{tabular}{lccc}
Matchers	&		& Open		& Close		\\
\hline
Parentheses	& \verb|()|	& \verb|[(<)]|	& \verb|[(>)]|	\\
Brackets	& \verb|[]|	& \verb|[[<]]|	& \verb|[[>]]|	\\
Braces		& \verb|{}|	& \verb|[{<}]|	& \verb|[{>}]|	\\
\end{tabular}
\end{center}

An experimental library and command-line tool
to parse MinML and convert it to HTML or XML,
written in Go,
is available at \url{https://github.com/dedis/matchertext}.
An \href{https://github.com/bford/hugo}{extention}
to the \href{https://gohugo.io}{Hugo}
static website generator supports web authoring in MinML.

%% file: eval.tex
\section{Evaluation}
\label{sec:eval}

This section is a placeholder at present,
pending more implementation and evaluation experience to report.

Some key questions we wish to evaluate include:
\begin{itemize}
\item	What are the most common kinds of embeddings
	that appear in large repositories of real source code,
	in what host and embedded language combinations,
	and for what purposes?
\item	How common and what kinds of needs are there for
	multiple levels of embedding in practice?
\item	For a set of popular (big or little) languages,
	how common are natural violations of matchertext discipline?
	Of what kinds are most common (e.g., in string constants, comments)?
	How painful would it be to fix these violations in typical code?
\item	How common and painful are needs to escape embedded strings
	when manually embedding into surrounding language strings?
	For example, how commonly do actual URIs embedded into
	actual program code need or use manual escaping?
\item	How common have security bugs been related to
	inadequate or incorrect escape armoring
	when embedding untrusted content automatically?
\item	What is the security-critical ``attack surface''
	(\eg code size and complexity)
	of typical string sanitizing mechanisms
	for embedding of untrusted content?
\item	What are the syntactic ``horror stories'' of cross-language embedding,
	akin to leaning toothpick syndrome,
	but perhaps in other combinations of languages
	and/or resulting in other symptoms?
\end{itemize}

\xxx{Prevalence of embedding versus unmatched matchers:

The key advantage of matchertext is to simplify embedding,
but it has the cost of making it more cumbersome
to express unmatched matchers.
Is this a worthwhile tradeoff?

\textbf{Experiment:}
For each of several popular programming languages,
(a) find a large repository of source code in that language
(\eg an open source software distribution);
(b) find a mature scanner/parser for that language;
(c) write a heuristic recognizer for a variety
of commonly-embedded strings in other ``big'' or ``little'' language syntaxes
(\eg regexes, URIs, IP addresses, JavaScript, $dots$);
(d) count the number of times that cross-language embeddings
appear in some form,
and the number of times that unmatched matchers appear.
Record and provide statistics on the contexts in which embeddings
and unmatched matchers appear
(\eg in function calls, if statements, case statements, comments, $dots$).

Priority languages to evaluate include:
HTML (especially detecting embedded JavaScript);
JavaScript/TypeScript (especially with embedded HTML);
C, C++, Java, Go, Swift, Perl, Raku, PHP.

It would probably be worth trying to
make the heuristic embedded string recognizer portable across languages
or find an existing one that is in some way.

}

%% file: rel.tex
\section{Related Work}
\label{sec:rel}

The theory of syntactic structures,
such as regular and
context-free languages~\cite{chomsky59algebraic}
or parsing expression grammars~\cite{ford04popl},
has already been richly developed.
Nothing about the matchertext discipline
is particularly new or technically challenging
from a formal language perspective.
However, surprisingly little prior work has focused on
the ubiquitous practice of synactically embedding
strings of one language into those of another,
or addressing the practical challenges this embedding creates.

Some recent work has focused on developing better tooling
to support string-embedding practices as they currently stand:
\eg parsing regular approximations
of string-embedded languages~\cite{verbitskaia15relaxed},
and support in
integrated development environments~\cite{grigorev14string}
and static analysis tools~\cite{khabibullin15development}.
This work does not attempt to explore syntax design practices
that could make languages more cleanly embeddable in the first place, however.

Significantly more work has focused on
\emph{domain specific embedded langauges}
or DSELs~\cite{hudak98modular},
particularly in the functional programming community.
DSELs build upon the syntax and semantics
of a general-purpose programming language such as Haskell,
and thus benefit from --
but also become dependent upon and specialized to --
the syntax, semantics, and tooling of the host language.
DSELs are thus unsuitable for embedded languages
that wish to remain agnostic to, or usable across a variety of,
host languages.
Asking an embedded language to conform only to the matchertext discipline --
that ASCII matchers must match --
is a much more lightweight proposition than
asking the embedded language to adopt, and become usable \emph{only} with,
the entirety of Haskell or another general-purpose programming language.

Pragmatically,
some existing languages come close to the matchertext approach to embedding.
Strings in PostScript~\cite{adobe99postscript}
are delimited by matching parentheses,
and may contain unescaped literal parentheses provided they are balanced.
The \verb|dc| calculator~\cite{howard21dc}
similarly delimits strings with brackets,
and allows balanced nested brackets.
\href{https://spec.commonmark.org/0.30/#links}{Link syntax}
in Markdown~\cite{macfarlane19commonmark},
like \verb|[|\emph{text}\verb|](|\emph{url}\verb|)|,
allows brackets within \emph{text}
and parentheses within \emph{url} provided they are balanced.
These languages use traditional backslash escapes to handle unbalanced matchers,
however,
and do not support general cross-language embedding
in the way matchertext does.

\xxx{ in this space, not sure if we should go into more detail:
\href{https://www.cambridge.org/core/services/aop-cambridge-core/content/view/4B0A7526CC16907F445CCF27277E9B9B/S0956796802004574a.pdf/div-class-title-compiling-embedded-languages-div.pdf}{Compiling embedded languages},
\href{https://ieeexplore.ieee.org/document/685738}{Modular domain specific languages and tools}
\href{https://scg.unibe.ch/archive/papers/Reng10aEmbeddingLanguages.pdf}{Embedding Languages Without Breaking Tools}
}

This related work is preliminary and no doubt incomplete;
proposals for relevant additions are welcome.

%% file: concl.tex
\section{Conclusion}
\label{sec:concl}

Matchertext is a syntactic discipline
that enables the verbatim embedding of strings across languages
with no transformation or expansion,
by enforcing only the single rule that the ASCII matchers --
parentheses, square brackets, and curly braces --
must always appear strictly in matched pairs.
We develop the basic principles and rationale for matchertext,
and explore potential backward-compatible extensions
to existing classes of languages
both to host embedded matchertext strings,
and to write embeddable matchertext conveniently.
Further experience is needed implementing and using matchertext
in order to evaluate its benefits and costs empirically.

\subsection*{Acknowledgments}

I wish to thank Terence Parr and Russ Cox
for valuable feedback on earlier drafts of this paper.

\xxx{samller contributions and funding acknowledgments here}